\documentclass[aps,prd,preprint,floatfix]{revtex4}
\usepackage{epsfig}
\usepackage{latexsym}

\begin{document}
 
\preprint{\rightline{ANL-HEP-CP-04-72}}
  
\title{Simulating lattice QCD at finite temperature and zero quark mass
\footnote{talk presented by D.~K.~Sinclair at the Workshop on QCD in Extreme
Environments, Argonne National Laboratory, 29th June to 3rd July, 2004.}}

\author{J.~B.~Kogut}
\address{Dept. of Physics, University of Illinois, 1110 West Green Street,
Urbana, IL 61801-3080, USA}
\author{D.~K.~Sinclair}
\address{HEP Division, Argonne National Laboratory, 9700 South Cass Avenue,
Argonne, IL 60439, USA}
  
\begin{abstract}
We simulate lattice QCD with an irrelevant chiral 4-fermion interaction which
allows us to simulate at zero quark mass. This enables us to study the 
finite-temperature chiral-symmetry-restoring phase transition for 2 massless
quark flavours, which is believed to be second order. In particular, it enables
us to estimate the critical exponents which characterize the universality
class of this transition. Our earlier simulations on $N_t=4$ and $N_t=6$
lattices revealed that finite lattice-spacing artifacts on such coarse lattices
affect the nature of the transition. We are now simulating on $N_t=8$ lattices
($12^3 \times 8$, $16^3 \times 8$ and $24^3 \times 8$ lattices) where we expect
to expose the continuum behaviour of this transition.
\end{abstract}

\maketitle

\section{Introduction} 

In the standard formulations of lattice QCD, the Dirac operator becomes singular
in the chiral (zero quark mass) limit. This prevents one from simulating at
zero quark mass, and makes simulations at small quark masses prohibitively
expensive. In addition, since the chiral condensate vanishes on a configuration
by configuration basis in this limit, even if one were able to perform zero-mass
simulations, such simulations would tell us nothing about spontaneous chiral
symmetry breaking.

We have modified the standard staggered-quark action for lattice QCD by the
addition of a chiral 4-fermion interaction of the form of Gross-Neveu chiral 
extension of the Nambu-Jona-Lasinio model 
\cite{Kogut:1998rg,Kogut:2000qp,Kogut:2001qu,Kogut:2002rw}. The idea of
combining gauge theories with 4-fermion interactions has been explored
by others \cite{Kondo:1988qd,Brower:1994sw,Brower:1995vf}.
We choose our 4-fermion operator to
preserve the reduced symmetries of the staggered-quark action.

This new action, by generating a chiral-symmetry preserving ``dynamical quark 
mass'', renders the Dirac operator non-singular, even for zero quark mass. In
addition, the chiral condensate also remains finite in the chiral limit on
each configuration. The chiral condensate vanishes as it must on any finite
lattice, because the orientation of the condensate rotates from configuration
to configuration such that the average over the infinite ensemble of 
configurations vanishes. This method of enforcing the absence of spontaneous
symmetry breaking on a finite lattice is identical to that seen in simple spin
models, giving us confidence that we should be able to use this action to
study critical behaviour of QCD, in the same way that these spin models allow
determination of their critical behaviour using Monte-Carlo methods (see for 
example \cite{Cucchieri:2002hu}).

Two-flavour QCD is expected to have a second-order finite temperature
chiral-symmetry-restoring (deconfining) phase transition at zero quark mass.
This transition is expected to lie in the universality class of the
3-dimensional $O(4)$ sigma model. With the reduced symmetry of the staggered
lattice we would expect this transition to lie in the universality class of
the $O(2)$ sigma model, until the lattice spacing is fine enough that the
transition is sensitive to the continuum symmetry. At non-zero quark mass,
this transition is expected to soften to a mere crossover. Hence to study this
second order transition, it is necessary to simulate at zero quark mass (or so
close to zero that the chiral extrapolation is unambiguous).

Our earlier simulations on $N_t=4$ \cite{Kogut:1998rg,Kogut:2002rw} 
and $N_t=6$ lattices \cite{Kogut:2000qp,Kogut:2001qu} showed the nature of
the transition to be sensitive to lattice artifacts on such coarse lattices.
The transition for $N_t=4$ could be made first or second order, depending on
the size of the 4-fermion coupling. The $N_t=6$ transition appeared to be
second order, but the measured critical exponents were those characteristic of
a tricritical point, rather than those of the $O(2)/O(4)$ spin models. We are
now extending these simulations to $N_t=8$ where the lattice artifacts will be
appreciably smaller.

In section~2 we define our new, ``$\chi$QCD'', staggered action. Section~3
describes our ongoing $N_t=8$ simulations and presents some preliminary
results. In section~4 we give our conclusions and plans for the future.

\section{Lattice QCD with chiral 4-fermion interactions}

The standard staggered quark action differs from the continuum action by terms
${\cal O}(a^2)$, where $a$ is the lattice spacing. This means we have the
freedom to add terms which only contribute at ${\cal O}(a^2)$ relative to the
standard action, i.e. dimension 6 or higher operators. All such additions
will do is change the size of the ${\cal O}(a^2)$ corrections and hence the
approach to the continuum. However, such actions will have the same continuum
limit as the standard action.

We choose to add a chiral 4-fermion interaction of the Gross-Neveu form 
\cite{Gross:1974jv,Nambu:1961tp,Nambu:1961fr} to
the standard staggered action. Since we are interested in 2-flavour QCD, we
choose an action whose naive continuum form would be
\begin{equation}
{\cal L}_f=\bar{\psi}(D\!\!\!\!/+m)\psi
        -{\lambda^2 \over 6 N_{2 \times 2}}[(\bar{\psi}\psi)^2
                          -(\bar{\psi}\gamma_5\tau_3\psi)^2].
\label{eqn:lagrangian2}
\end{equation} 
where $N_{2 \times 2}$ is the number of $SU(2) \times SU(2)$ flavour doublets.
Since the fermion doubling in the staggered fermion lattice implementation
leads naturally to 4 flavours, the actual continuum action will be
\begin{equation}                                                              
{\cal L}_f=\bar{\psi}(D\!\!\!\!/+m)\psi                                       
        -{\lambda^2 \over 6 N_{4 \times 4}}[(\bar{\psi}\psi)^2                
                          -(\bar{\psi}\gamma_5\xi_5\psi)^2]. 
\label{eqn:lagrangian4}                                        
\end{equation}
where $N_{4 \times 4}$ is the number of $SU(4) \times SU(4)$ flavour quartets.
Clearly these two forms are equivalent if we replace 
$\lambda^2 \rightarrow \lambda^2/2$ since $N_{4 \times 4}=N_{2 \times 2}/2$.

Now we turn to the staggered lattice implementation of QCD with such 4-fermion
interactions \cite{Aoki:1991px}. The fermion part of the action is
\begin{equation}
S_f=\sum_{sites}\bar{\chi}[\not\!\! D + m +
\frac{1}{16} \sum_i (\sigma_i+i\epsilon\pi_i)]\chi
+\sum_{\tilde{s}}\frac{1}{8}N_f\gamma(\sigma^2+\pi^2),
\end{equation}
where $\epsilon(x)=(-1)^{x+y+z+t}$, $\sigma$ and $\pi$ are auxiliary fields
residing on the dual lattice and $i$ runs over the 16 sites on the dual lattice
closest to the site $x$. Integrating out these auxiliary fields gives the
lattice version of equation~\ref{eqn:lagrangian4}, remembering that the total
number of flavours $N_f=4 N_{4 \times 4}$. Here $\gamma=3/\lambda^2$. Since
the fermion determinant becomes real in the continuum limit, we use the
determinant of $M^{\dagger}M$ for our simulations and take the 4th root of this
determinant using the hybrid molecular-dynamics method with `noisy' fermions
to get $N_f=2$. We refer to lattice QCD with this action ``$\chi$QCD''.

It is easy to see that this action preserves the same $U(1)$ vestige of the
$SU(4) \times SU(4)$ continuum chiral symmetry that is preserved by the
standard staggered action at $m=0$, which is why this particular form was 
chosen.

The reason it is possible to simulate this action at zero quark mass is clear
in the confined phase. Here $\bar{\psi}\psi$ develops an expectation value.
Hence since
\begin{equation}
\langle\sigma\rangle = \langle\bar{\psi}\psi\rangle/\gamma
\end{equation}
the quark field develops an effective mass $m=\langle\sigma\rangle$, which
makes the Dirac operator non-singular. What is less clear is why the Dirac
operator remains non-singular in the chirally symmetric phase.

$\gamma$ should be chosen large so that the 4-fermion coupling is small and thus
any additional flavour symmetry breaking over that already present in the 
standard staggered action, is also small. In particular $\gamma$ should be
chosen far larger than that for which there is a bulk transition in the absence
of the gauge fields. In addition we shall try to keep $\gamma$ large enough
that the critical value of $\beta=6/g^2$ is not too far from that without the
4-fermion term, i.e. the gauge interactions should be primarily responsible
for quark binding and chiral symmetry breaking.

\section{``$\chi$QCD'' at finite temperature.}

Since $N_t=4$ and $N_t=6$ simulations showed evidence that the nature of the
finite temperature phase transition was being affected by finite lattice
spacing artifacts, we are now simulating at $N_t=8$. In particular we are
running on $12^3 \times 8$, $16^3 \times 8$ and $24^3 \times 8$ lattices.

We are currently running at zero quark mass with $\gamma=10$. At this $\gamma$
we are able to use a molecular-dynamics time increment of $dt=0.05$, and find
that at the worst it takes 600-700 conjugate gradient iterations to invert the
Dirac operator. However, because we now have a true critical point in the
infinite volume limit, we have to face critical slowing down. For this reason
we ran for 50,000 time units for each $\beta$ in the transition region and are
extending these runs to 100,000 time units. 

\begin{figure}[htb]    
\epsfxsize=6in     
\centerline{\epsffile{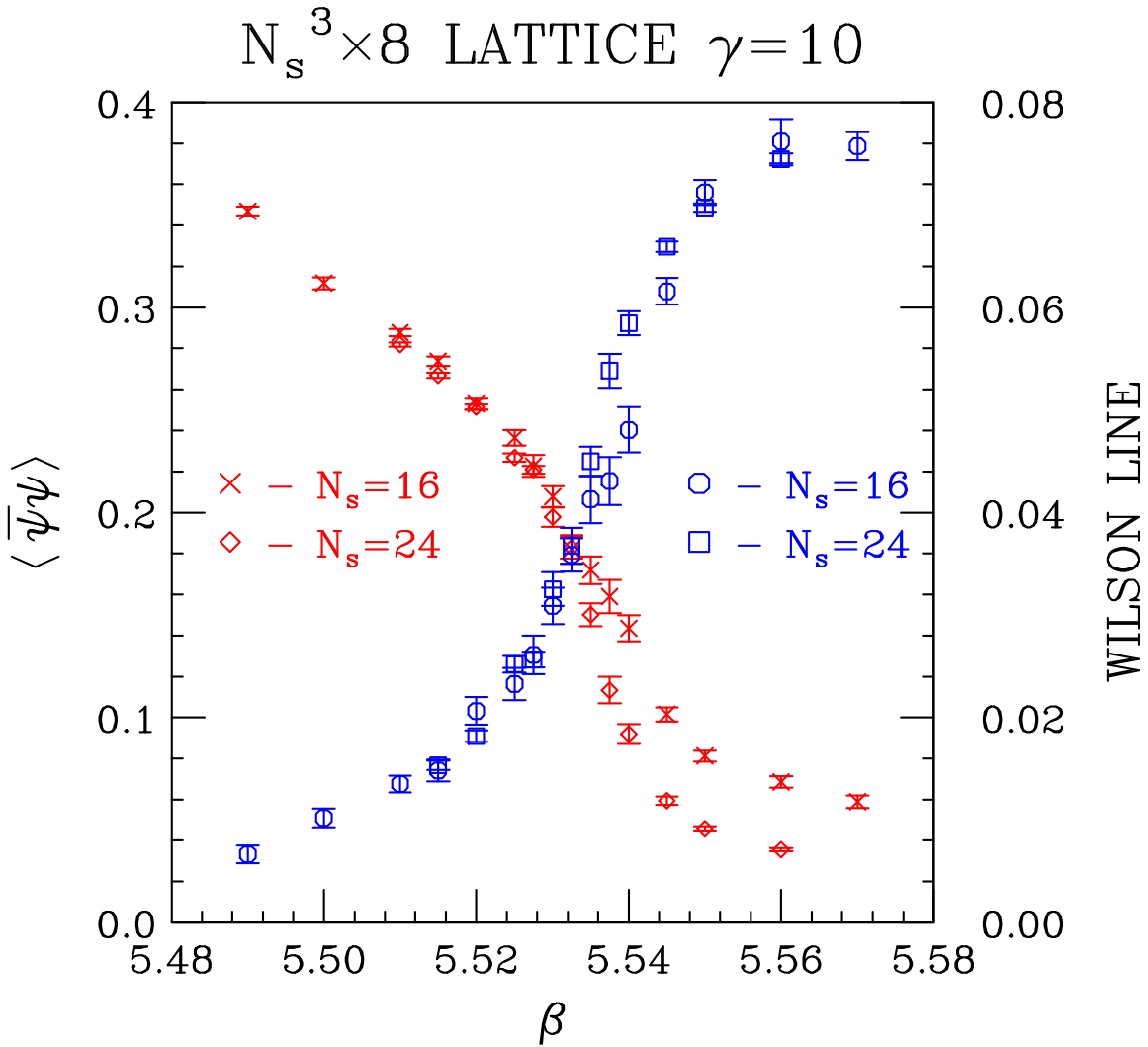}}
\caption{Wilson Line and chiral condensates as functions of $\beta$.}
\label{fig:wil-psi}
\end{figure}

Let us now look at our preliminary results. In figure~\ref{fig:wil-psi} we
show the Wilson Line (Polyakov Loop) and the chiral condensate as functions of
$\beta$ across the transition region. The chiral condensate plotted here is
really $\sqrt{\langle\bar{\psi}\psi\rangle^2-
\langle\bar{\psi}\gamma_5\xi_5\psi\rangle^2}$.
What we see is that the transition region, 
$5.530 \lesssim \beta \lesssim 5.545$, is very narrow, which explains why it
is so difficult to make quantitative measurements for standard actions at
non-zero quark masses. The crossover appears quite smooth which suggests that
it is second order. Long-time correlations, which are apparent when one looks
at the time histories of the measured observables close to the transition, also
point to the transition being second order.

\begin{figure}[htb]
\epsfxsize=3in
\epsffile{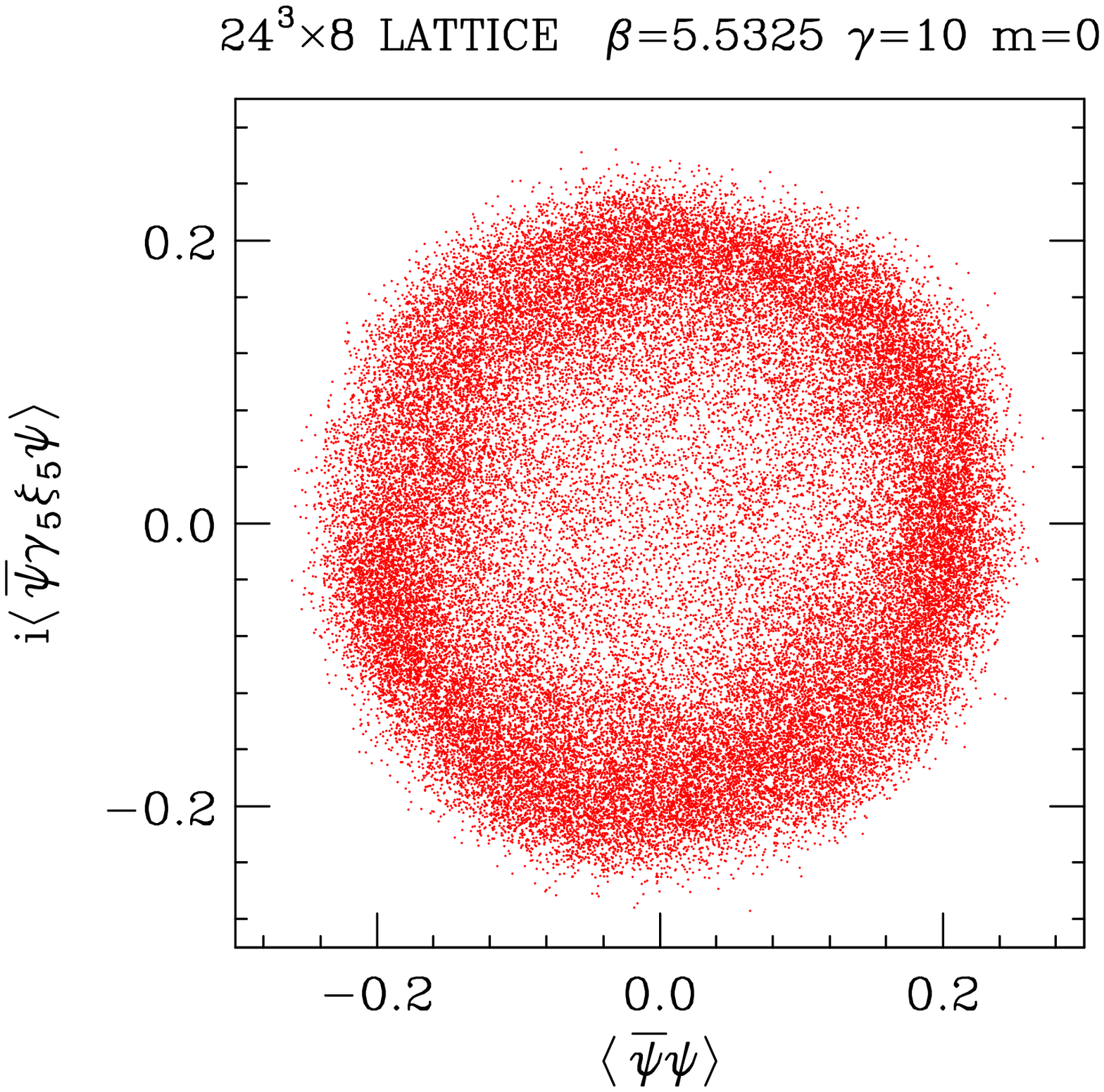}
\epsfxsize=3in
\epsffile{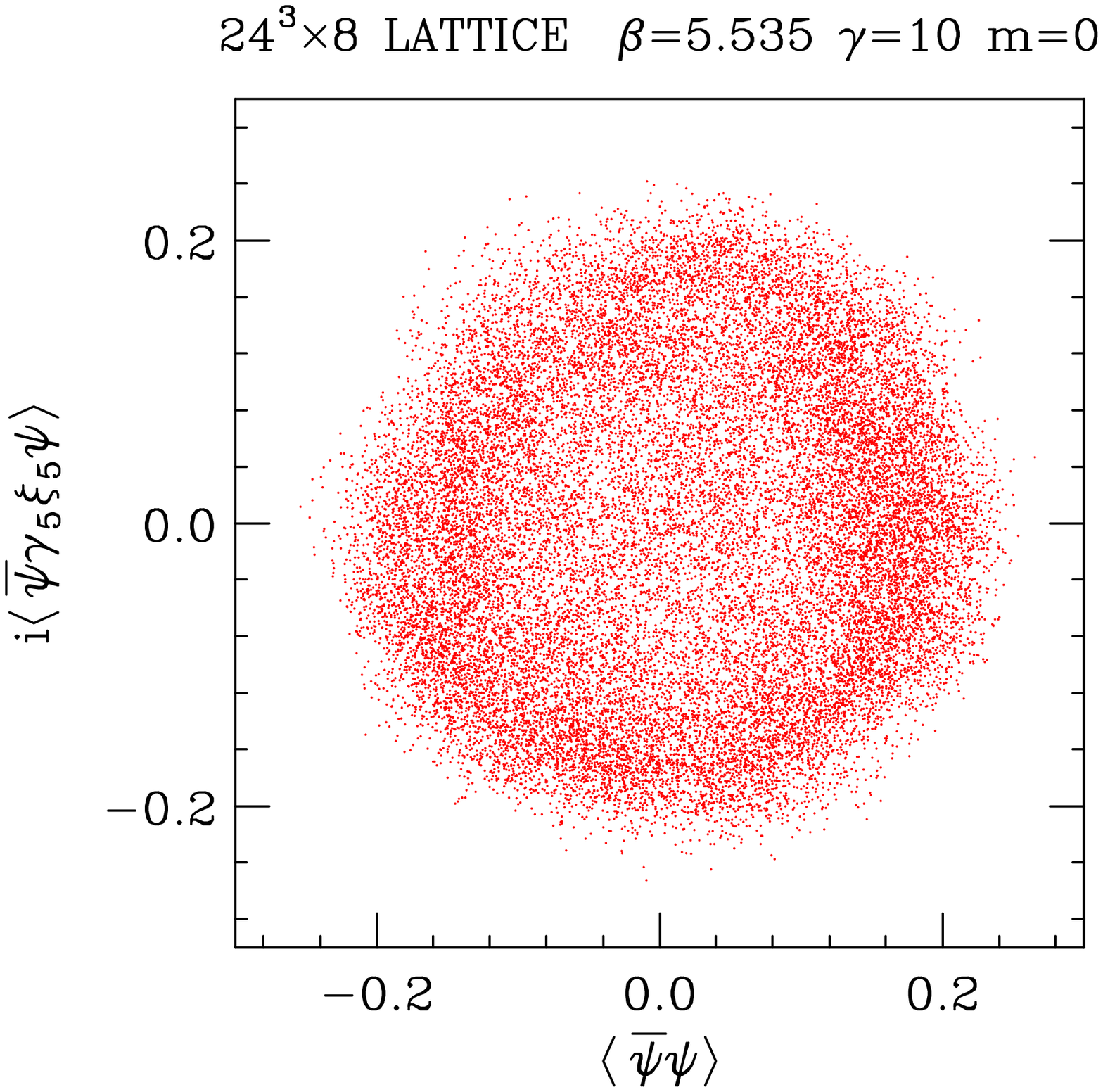}
\epsfxsize=3in
\centerline{\epsffile{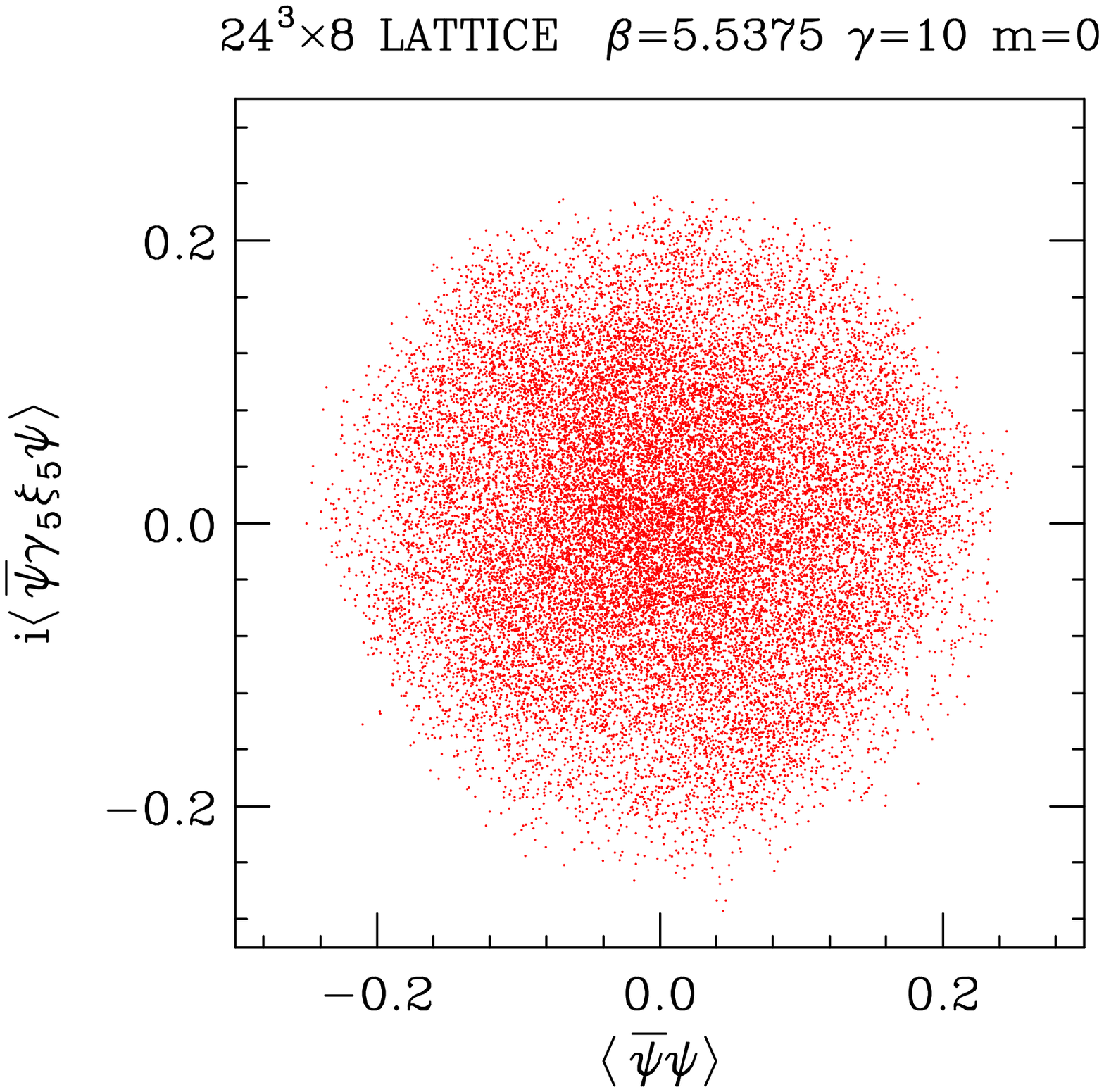}}
\caption{Distribution of chiral condensates for a) $\beta=5.5325$,
         b) $\beta=5.535$ and c) $\beta=5.5375$.}
\label{fig:chiral}
\end{figure}

\begin{figure}[htb]
\epsfxsize=3in
\epsffile{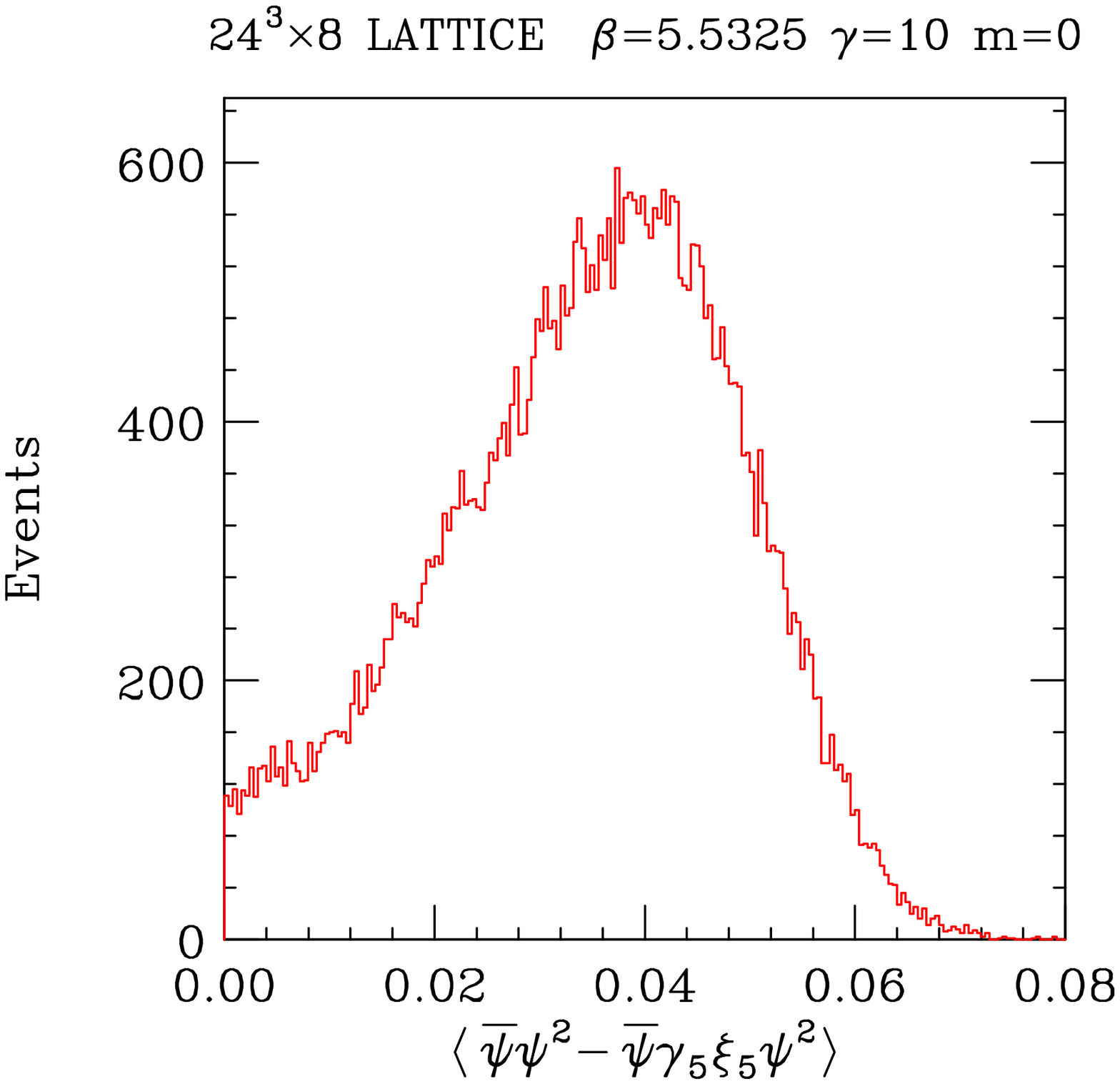}
\epsfxsize=3in   
\epsffile{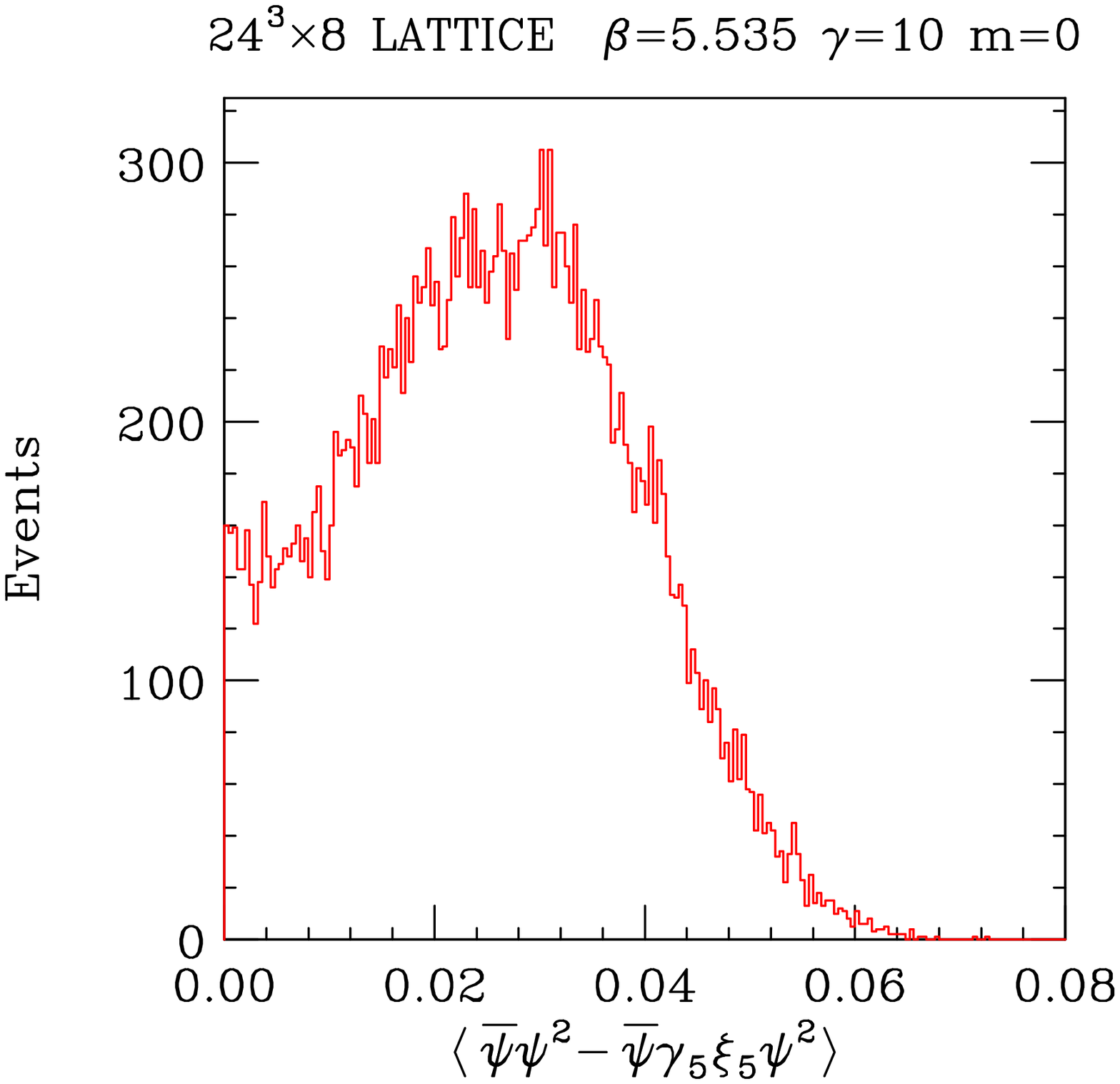}
\epsfxsize=3in                    
\centerline{\epsffile{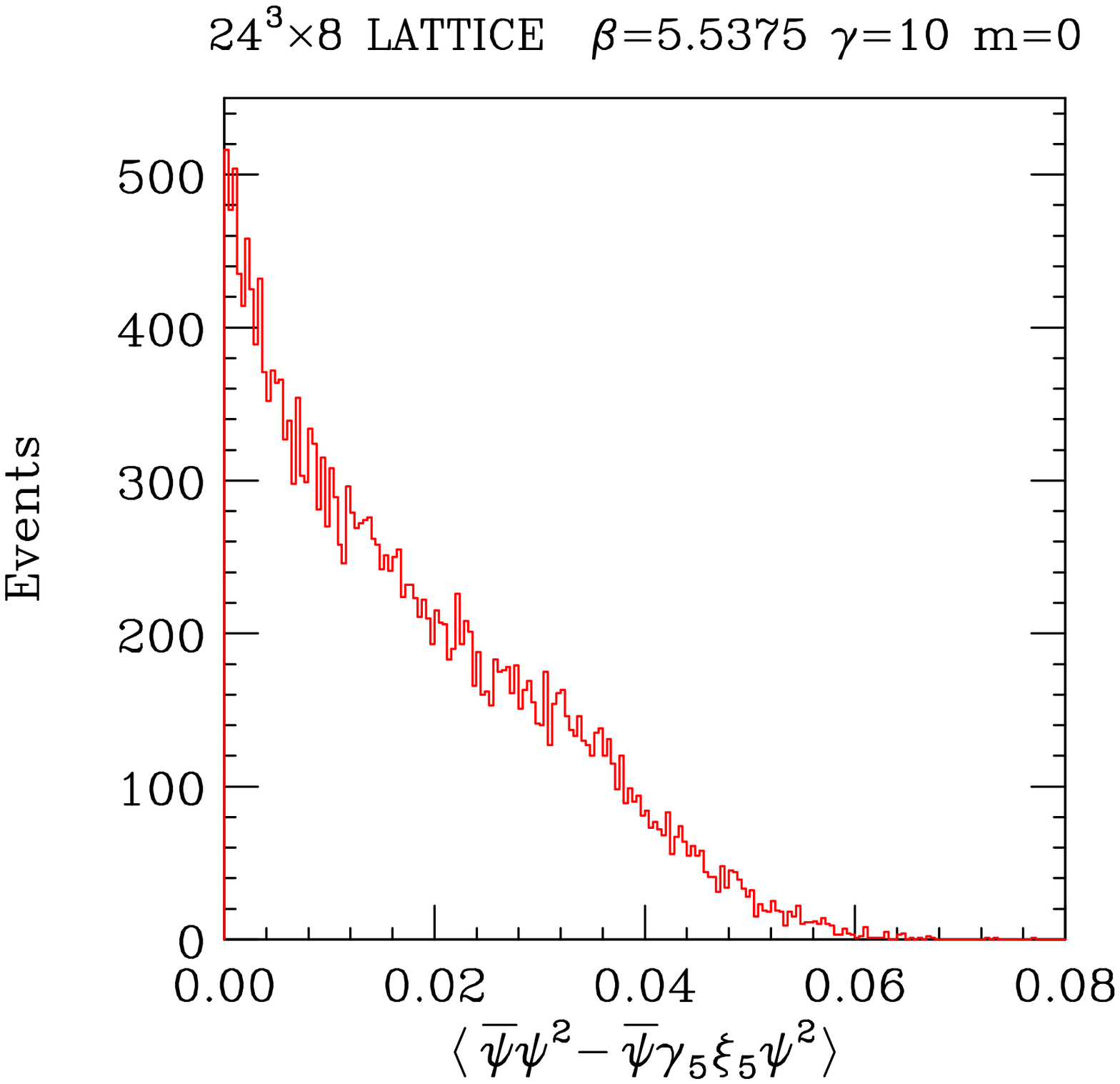}} 
\caption{Histograms of distributions of chiral condensates for 
a) $\beta=5.5325$, b) $\beta=5.535$ and c) $\beta=5.5375$.}  
\label{fig:chiral_hist}                                                       
\end{figure}

The chiral order parameter `$\langle\bar{\psi}\psi\rangle$' or 
`$\langle\sigma\rangle$' is non-zero on each configuration, but its orientation
in the $(\langle\bar{\psi}\psi\rangle,
i\langle\bar{\psi}\gamma_5\xi_5\psi\rangle)$ or equivalently 
$(\langle\sigma\rangle,\langle\pi\rangle)$ plane rotates from configuration
to configuration (here the $\langle\rangle$ represents the lattice average
for a given configuration). It is this rotation that makes the ensemble
average zero as it must be for a finite lattice. In figure~\ref{fig:chiral},
we show the distribution of values of the chiral condensate during runs at
several $\beta$s on a $24^3 \times 8$ lattice. What we notice for
$\beta=5.5325$ and $\beta=5.535$ is that these points are concentrated in an
annulus of non-zero radius. This is an indication that these $\beta$s lie in
the low-temperature phase where the chiral order parameter is non-zero. For
$\beta=5.5375$ there is no such annulus and the distribution peaks near the
origin. This suggests that $\beta=5.5375$ lies in the high temperature phase,
where the chiral order parameter is zero and chiral symmetry is restored. From
these observations, we would predict that the critical $\beta$, $\beta_c$ lies
in the range $5.5350 \lesssim \beta_c \lesssim 5.5375$.
We also note that these distributions appear to have a circular symmetry in
the plane. This ensures that the ensemble average will be zero. To see the
distribution in the radial coordinate more clearly we have histogrammed these
distributions to equal-size bins in 
$\langle\bar{\psi}\psi\rangle^2-\langle\bar{\psi}\gamma_5\xi_5\psi\rangle^2$
which yields a quantity proportional to the density in the planes of 
figure~\ref{fig:chiral}. These histograms are presented in 
figure~\ref{fig:chiral_hist}.

\begin{figure}[htb]
\epsfxsize=6in
\centerline{\epsffile{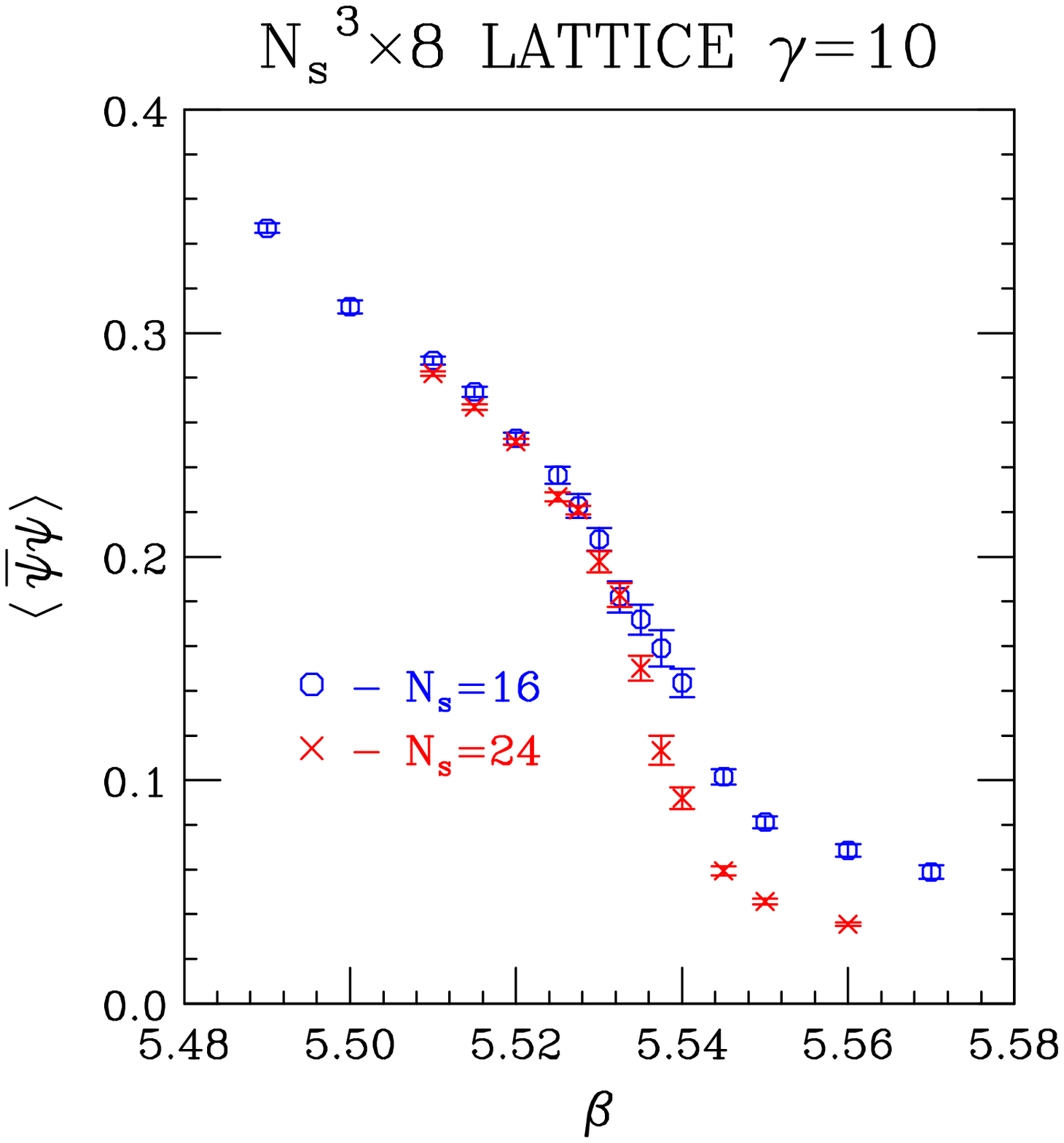}}
\caption{Chiral condensates as functions of $\beta$.}
\label{fig:pbp}
\end{figure}

Now let us examine the chiral condensate in more detail. In 
figure~\ref{fig:pbp} we plot the chiral condensates from 
figure~\ref{fig:wil-psi}. 
Because of the way in which the chiral condensate rotates from configuration
to configuration, any methods used to derive the ensemble averaged condensate
which has the correct continuum limit in the ordered phase will, of necessity,
smooth out the transition. It is then a matter of `experimentation' to determine
which method is easiest to extrapolate to the infinite volume limit. The
figure shows the average of the magnitude of the condensate, which has enjoyed
some success with analyzing spin models \cite{Cucchieri:2002hu,Talapov:1996yh}.
The other method which we have used is to average the squares of the 
condensates. 

To determine the first of the critical exponents $\beta_m$, we need to fit
the chiral condensate to the scaling form
\begin{equation}
\langle\bar{\psi}\psi\rangle = c (\beta - \beta_c)^{\beta_m}
\end{equation}
for $\beta$ close to $\beta_c$. $\beta_m$ serves to distinguish the universality
class of the transition. $\beta_m=0.384(5)$ for $O(4)$, $\beta_m=0.35(1)$ for
$O(2)$, $\beta_m=1/4$ for a tricritical point and $\beta_m=1/2$ for a
mean-field transition 
\cite{baker,LeGuillou:1979cf,Kanaya:1994qe,Butera:1995kr,lawrie}.
While it is doubtful if we will accumulate enough
statistics to distinguish $O(2)$ from $O(4)$ behaviour, we should be able to
distinguish $O(2)/O(4)$ behaviour from mean-field or tricritical behaviour.

Fitting our current $24^3 \times 8$ `data' to this scaling form over the range
$5.5275 \le \beta \le 5.5375$ gives $\beta_m=0.4(1)$. This should improve when
we finish extending our statistics from 50,000 to 100,000 trajectories for all
$\beta$s in the scaling window. In addition, we have yet to extract the
maximum information from our `data' by using Ferrenberg-Swendsen interpolation
methods \cite{Ferrenberg:yz} 
to increase the number of $\beta$ values contributing to our fit.  

Another method to obtain information on the position and nature of the 
transition is to use 4th-order Binder cumulants for our order parameter, for
several lattice sizes. The 4-th order cumulant for the auxiliary fields is
defined as \cite{binder}
\begin{equation}
B_4 = { \langle(\vec{\sigma}^2)^2\rangle \over 
        \langle(\vec{\sigma}^2)\rangle^2 }
\end{equation}
where $\vec{\sigma}=(\sigma,\pi)$. Here we use the auxiliary fields rather than
$(\langle\bar{\psi}\psi\rangle,i\langle\bar{\psi}\gamma_5\xi_5\psi\rangle)$
because we only have a single stochastic estimator for these fermionic
order parameters for each trajectory, and we would need at least 4 such 
estimators/configuration to be able to obtain an unbiased estimator of $B_4$.
Note that $B_4$ uses quantities measured on finite lattices, and does not
suffer from the difficulties one encounters with trying to estimate the chiral
condensate itself.

\begin{figure}[htb]
\epsfxsize=6in
\centerline{\epsffile{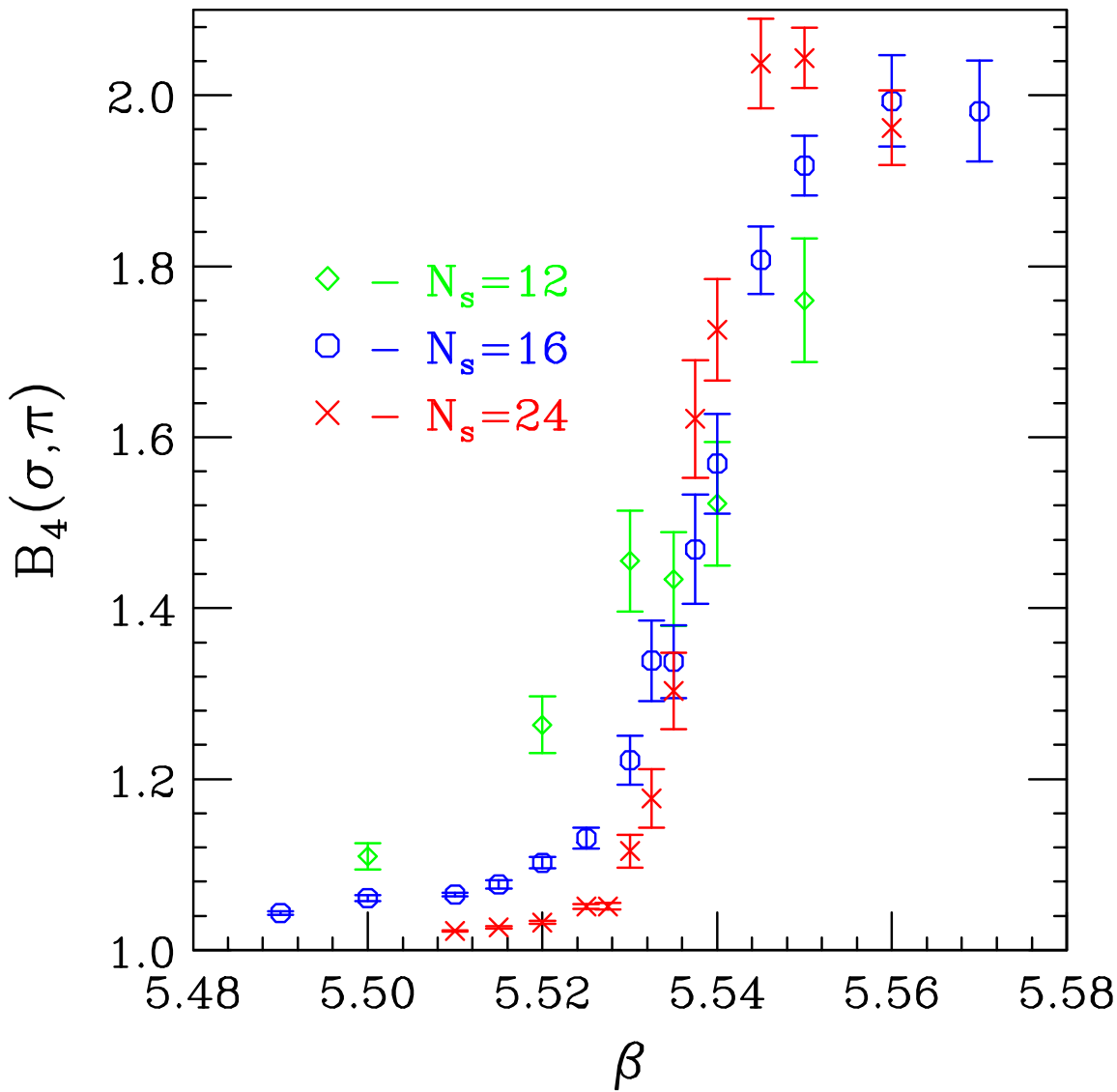}}
\caption{Binder cumulants as functions of $\beta$.}
\label{fig:binder}
\end{figure}

$B_4=1.242(2)$ for $O(2)$ \cite{Cucchieri:2002hu}. $B_4=1.460998...$ 
for a tricritical point and $B_4=\pi/2$ for a mean-field transition 
\cite{lipowski}. $B_4=2$ for a crossover. $B_4=1$ for a first-order transition.
These are all infinite volume results. While the $B_4$ values associated with
a crossover or a first-order transition are approached slowly with increasing
volume close to the second-order transition, the infinite volume value is
approached rapidly with increasing volume, at the transition. As $\beta
\rightarrow 0$, $B_4$ approaches its first-order value, while as $\beta
\rightarrow \infty$ $B_4$ approaches its crossover value. The rate at which
these values are approached as one goes away from the transition increases
with increasing volume. The curves for different volumes will cross very close
to the phase transition, yielding both estimates for $B_4$ and hence of the
universality class of this critical point, and of $\beta_c$ for this
transition. Our preliminary results for all 3 lattice sizes are given in
figure~\ref{fig:binder}.
Although these measurements are very noisy, the curves for the 3 lattice sizes
appear to cross at $5.5325 \lesssim \beta_c \lesssim 5.5375$. The value at
which they cross is uncertain enough to accommodate both $O(2)$ and tricritical
behaviour.

Perhaps the best way to estimate the position of the phase transition is from
the peaks of the susceptibilities. For any observable $X$ the corresponding
susceptibility is defined by
\begin{equation}
\chi_{_{\scriptstyle X}} = V \langle (X -\langle X \rangle)^2 \rangle.
\end{equation}
We use Ferrenberg-Swendsen interpolation to locate the peak of the 
susceptibilities for the plaquette, the Wilson line and $\vec{\sigma}$.
The advantage of this for determining $\beta_c$ is that these peaks give
estimates for the position of the transition on a finite volume, where it is
merely a crossover. These $\beta_c$ estimates are relatively insensitive to 
the lattice volume, which will make any attempt to extrapolate to infinite
volume somewhat easier. For the $24^3 \times 8$ lattice we estimate
$\beta_c=5.535(1)$ while for the $16^3 \times 8$ lattice we get
$\beta_c=5.537(2)$, by this method.

\section{Conclusions}

Lattice QCD, where the fermion action is modified to include a chiral 4-fermion
term shows promise for determining the precise nature of the $N_f=2$ finite
temperature phase transition for massless quarks, by determining the critical 
exponents of this transition. The ability to work at and near $m=0$ seems
essential for extracting this critical behaviour. With conventional actions,
the penalties for running at quark masses small enough to predict the chiral
limit appear insurmountable, at least in the near future. To see this one needs
only to look at the best `data' currently available from simulations using
state-of-the-art actions which do not permit using $m=0$ \cite{Bernard:2004je}. 
We have already seen that simulations with $N_t=4,6$ lattices are plagued by
finite lattice artifacts as was implicit in the analyses using standard
actions. We are therefore using lattices with $N_t=8$.

Preliminary fits to the $24^3 \times 8$ data yield critical exponent
$\beta_m=0.4(1)$, consistent with the expected $O(2)$ or $O(4)$ values. 
However, with an error this large, one cannot preclude tricritical or mean-field
behaviour. More statistics are needed as well as reliable extrapolations to
infinite volume. Use of Ferrenberg-Swendsen interpolation, thus making more
use of the limited `data' we do have, is one way of effectively increasing our
statistics.

The scaling window is very narrow $\Delta\beta \approx 0.01$, which is the
principal reason why extrapolating from finite quark mass is extremely
difficult. We are able to pinpoint the critical point from the peaks in the
susceptibilities for the various observables. This method yields the best
estimate of this position for a given lattice size and does not rely on some
potentially unreliable infinite volume extrapolation. Since the volume
dependence of $\beta_c$ determined this way is minimal, this is probably also 
the best method for determining $\beta_c$ in the infinite volume limit.

Binder cumulants show promise for determining the position and nature of the
phase transition. However, these require much more statistics than we have
obtained and possibly more than we can hope to obtain. They have the advantage
that they make use of the chiral condensate as measured on the finite lattice,
despite the fact that it averages to zero at all $\beta$s. Such are the
advantages of fluctuation quantities.

We will ultimately measure other critical exponents. One such, namely $\delta$,
measures the mass dependence of the condensate at $\beta_c$. Hence this will
require simulations at small but finite quark mass.

We are currently performing simulations with this action on zero temperature
lattices. These simulations will calculate the spectrum of light hadrons, which
should indicate the utility of this action for problems other than QCD
thermodynamics, for which the ability to simulate at zero or physical quark
masses is also important.

\section*{Acknowledgements}
JBK was supported in part by NSF grant NSF PHY03-04252. DKS is supported by DOE
contract W-31-109-ENG-38. These simulations are being performed on the IBM SP,
Seaborg at NERSC, on the IBM SP, BlueHorizon and the DataStar at NPACI and on
the Linux cluster, Tungsten at NCSA. Access to the NSF machines is provided
through an NRAC allocation.

\end{document}